\documentstyle[preprint,aps,tighten,epsf,floats]{revtex}
\newcommand{\ben}{\begin{displaymath}}
\newcommand{\een}{\end{displaymath}}
\newcommand{\be}{\begin{equation}}
\newcommand{\ee}{\end{equation}}
\newcommand{\bea}{\begin{eqnarray}}
\newcommand{\eea}{\end{eqnarray}}

\newcommand{\fign}[1]{\label{#1}}

\begin{document}
\draft
\title{Nucleon-Nucleon Scattering and Effective Field Theory: Including Pions
Non-perturbatively}
\author{ J. Gegelia}
\address{Department of Physics, The Flinders University of South Australia, \\
Bedford Park, SA 5042, Australia}
\date{\today}
\maketitle 

\begin{abstract}
Next to leading order effective field theory calculations are performed for $
{ }^1S_0$ $NN$ scattering using subtractive renormalization procedure. One
pion exchange and contact interaction potentials are iterated using
Lippman-Schwinger equation. Satisfactory fit to the Nijmegen data is obtained
for the momenta up to 300 MeV in the centre of mass frame. Phase shifts are also
compared with the results of KSW approach where pions are included
perturbatively.  
\end{abstract}


\section{Introduction}

There has been much recent interest in applications of 
the chiral perturbation theory approach to
processes involving an arbitrary number of nucleons \cite{all} introduced by
Weinberg \cite{weinberg1,weinberg2}.

For processes involving more than one nucleon Weinberg's power counting is
applied to the potential rather than to the scattering amplitude. For
$n$-nucleon processes the
potential is defined as a sum of $n$-nucleon irreducible time-ordered
perturbation theory diagrams. 
The amplitude is obtained by solving Lippmann-Schwinger equation (or
Schr\" odinger equation).

Iteration of the potential via
the Lippmann-Schwinger equation leads to divergences. One could try to
regularize the potential and include counter-terms, but due to the
non-renormalizability of the
theory one would have to include an infinite number of (counter-)terms with
more and more derivatives. 
Hence, one has either to exactly solve the equation and after subtract
divergences explicitly, or otherwise draw all relevant diagrams,
subtract them and sum up. There
are not any equations for subtractively renormalised amplitudes. 
Fortunately there exist practically more powerful approach, cut-off
EFT \cite{lepage}. Divergences appearing in diagrams can be 
regulated using (sharp or smooth) cut-off regularization. One can keep cut-off
parameter of the order of the mass of lightest integrated particle and fit
coupling constants to 
the experimental data as was done in ref. \cite{ordonez}. Cut-off EFT is
equivalent to subtractively renormalised EFT up to the order of considered
accuracy \cite{lepage,lepage1,gegelia2}. 
 
Technically more convenient approach based on a new systematic power counting
has been suggested by Kaplan, Savage and Wise
(KSW counting) in \cite{kaplan2}. Pions are included perturbatively within this
new scheme. A similar power counting with perturbative
pions has been suggested by Lutz \cite{lutz}. As was mentioned in \cite{kaplan3}
the 
expansion parameter in KSW counting is rather large. It raises a question of
the usefulness of suggested expansion. Arguments supporting as well as
criticising the perturbative inclusion of pions have been given in the
literature (see \cite{kapste} and citations therein. \cite{kapste} also contains
references to lots of papers devoted to EFT study of $NN$ interaction problem).

In Weinberg's power counting the one pion exchange potential is of leading order
and hence it has to be iterated via the Lippmann-Schwinger equation. 
KSW counting suggests that contributions containing multiple iterations of the
one-pion exchange potential are higher order and hence need not be included in
low order calculations. 
 
In this letter the results of the application of subtractive
renormalization procedure to the Weinberg's approach to ${ }^1S_0$
$NN$ scattering problem are briefly discussed (the details will be given in
separate paper).  Leading order (contact
interaction plus 
one-pion exchange) plus next-to-leading order contact interaction potential is
iterated using Lippman-Schwinger equation.


According Weinberg's power counting the leading order potential for ${ }^1S_0$
$NN$ scattering is the following:  

\be
V_0({\bf p},{\bf p'})=\tilde C+V_\pi ({\bf p},{\bf p'})
\label{potential}
\ee
where
\be
\tilde C\equiv C+{g_A^2\over 2f_\pi^2}, \ \ 
V_\pi ({\bf p},{\bf p'})\equiv -{4\pi\alpha_\pi\over \left( {\bf q}^2+
m_\pi^2\right)}, \ \ \alpha_\pi\equiv{g_A^2m_\pi^2\over 8\pi f_\pi^2}.
\label{potspel}
\ee
$C$ is determined by contact interaction terms in the effective Lagrangian
\cite{weinberg2}, ${\bf q}={\bf p}-{\bf p'}$, $g_A=1.25$ is the axial coupling
constant, $m_\pi =140 \ {\rm MeV}$ is the pion mass and $f_\pi =132 \ {\rm MeV}$
is the pion decay constant. 

Substituting $V_0$ into the Lippman-Schwinger equation one finds for the Feynman
amplitude \cite{kaplan}:

\be
iA=iA_\pi-i{\tilde C\left[ \chi_{\bf p}\right]^2\over
1-\tilde CG_E}
\label{amplitude}
\ee
where the quantities $A_\pi$, $\chi_{\bf p}$ and $G_E$ correspond to diagrams
drawn in FIG. 1. $A_\pi$ and  $\chi_{\bf p}$ are finite. The first diagram
contributing to $G_E$ is linearly and the second logarithmically divergent, all
other diagrams are finite. As $A$ contains divergences it is necessary to
regularize and renormalize it. 

Let us introduce a cut-off into the potential:
\be
V_0^{\Lambda}({\bf p},{\bf p'})={\tilde C \Lambda^8\over \left( {\bf
p}^2+\Lambda^2\right)^2\left( {\bf p'}^2+\Lambda^2\right)^2} 
+V_\pi ({\bf p},{\bf p'})
\label{cutpotential}
\ee
where $\Lambda $ is a cut-off parameter.

\begin{figure}[t]
\hspace*{2cm}  \epsfxsize=12cm\epsfbox{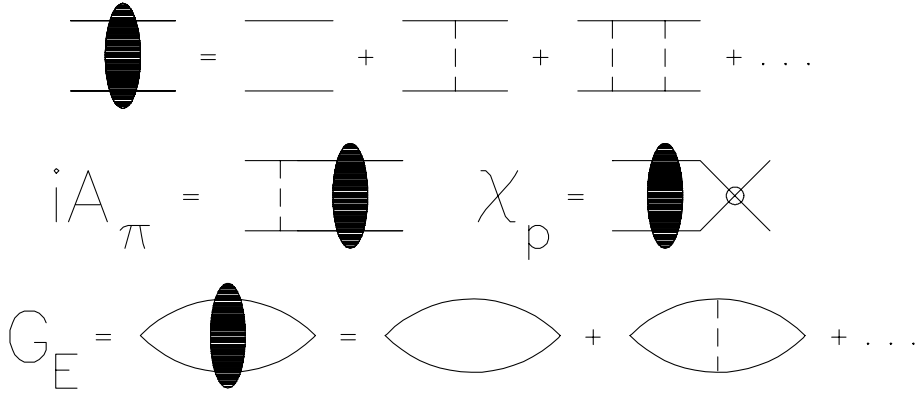}
\vspace{10mm}
\caption{\fign{apgexi}{\it 
Diagrams defining $A_\pi$, $\chi_{\bf p}$ and $G_E$. Solid and dashed lines
correspond to nucleon and pion respectively. Small circle in the diagram
defining 
$\chi_{\bf p}$ indicates that factor $\tilde C$ is excluded from that vertex.}}
\end{figure}

\begin{figure}[t]
\hspace*{3cm}  \epsfxsize=10cm\epsfbox{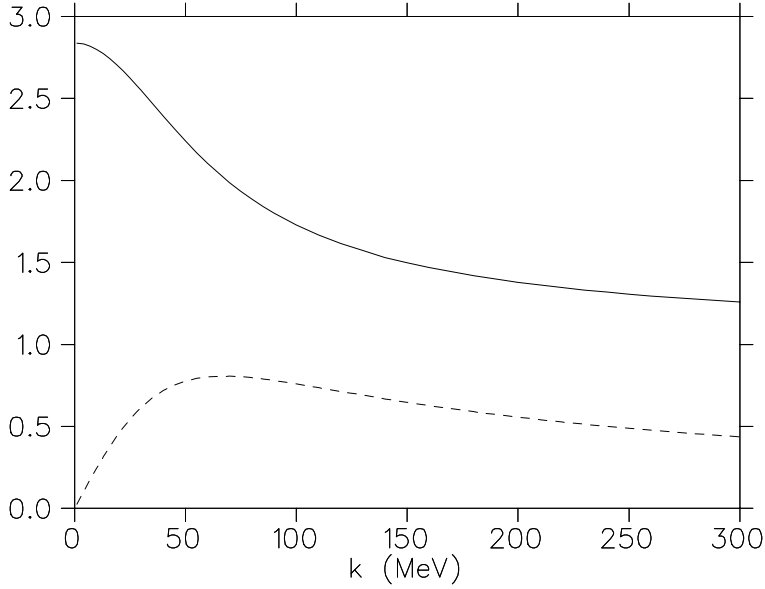}
\vspace{2mm}
\caption{\fign{xi}{\it 
$\left[ \chi_{\bf p}({\bf 0})\right]^2$ plotted versus momentum in {\rm
MeV}. The solid and dashed lines correspond to the real and imaginary parts
respectively.}} 
\end{figure}

\begin{figure}[t]
\hspace*{3cm}  \epsfxsize=10cm\epsfbox{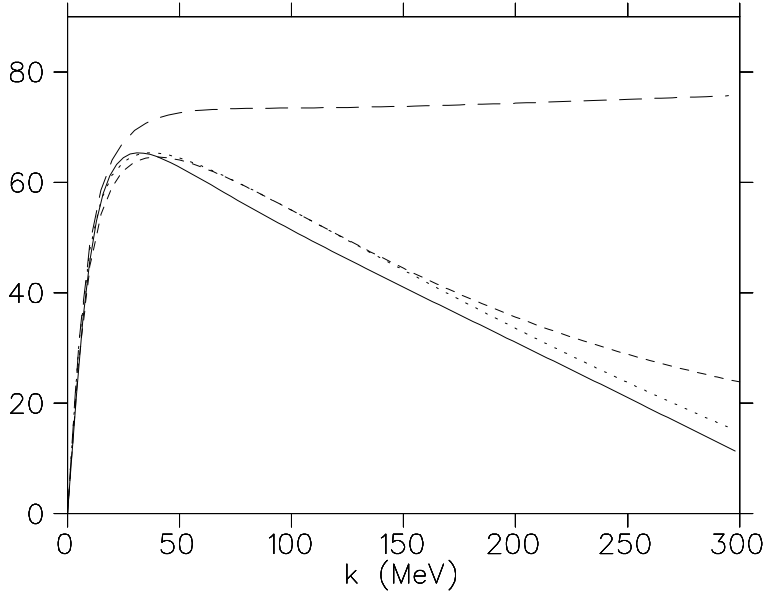}
\vspace{2mm}
\caption{\fign{npfig}{\it 
Phase shifts in degrees versus centre of mass momentum. Solid line corresponds
to Nijmegen phase shift data. Long-dashed line is the leading
order EFT result. Dotted and short-dashed lines are the results of 
next-to-leading order calculations. The dotted and small-dashed lines correspond
to non-perturbative and non-perturbative inclusion of next-to-leading order
potential $C_2\left( {\bf p}^2+{\bf p'}^2\right)/2$.}}
\end{figure}

Substituting $V_0^\Lambda$ into the Lippmann-Schwinger equation one gets the
following expression for the amplitude:
\be
iA^\Lambda =iA_\pi-i{\tilde C\left[ \chi_{\bf p}^\Lambda \right]^2\over
1-\tilde CG_E^\Lambda}
\label{cutamplitude}
\ee
where $A_\pi$ is unchanged, $\chi_{\bf p}^\Lambda $ and $G_E^\Lambda $ are given
by the 
same diagrams in FIG.1 with $\tilde C$ replaced by ${\tilde C \Lambda^8\over 
\left( {\bf
p}^2+\Lambda^2\right)^2\left( {\bf p'}^2+\Lambda^2\right)^2}$.

Note that $\chi_{\bf p}^\Lambda=\chi_{\bf p}+0(1/\Lambda )$ and 
$G_E^\Lambda =a_1\Lambda+a_2\ln \Lambda +G_E^f+0(1/\Lambda )$, where $G_E^f$ is
finite and $\Lambda $-independent.
It is straightforward to find $a_1$ and $a_2$ by calculating analytically first
two diagrams contributing to $G_E^\Lambda $. 

One can numerically calculate $A_\pi $ and $A^\Lambda $ for different values of
$\Lambda $. Using these values of $A_\pi $ and $A^\Lambda $ one finds: 
\be
{1\over A^\Lambda -A_\pi }=B_1(p)\Lambda +B_2(p)\ln\Lambda
+B_3(p).  
\label{A-A}
\ee
Although  $B_3(p)$ actually depends on $\Lambda $, for very large $\Lambda $ it
practically does not and $B_1(p)$ and $B_2(p)$ can be calculated very
accurately. Having calculated $a_1$, $a_2$, $B_1(p)$ and $B_2(p)$ one easily
finds:
\be
\left[ \chi_{\bf p}^\Lambda\right]^2={a_1\over B_1(p)}
\label{chi1}
\ee

\be
\left[ \chi_{\bf p}^\Lambda\right]^2={a_2\over B_2(p)}
\label{chi2}
\ee
Good agreement between numerical results obtained from (\ref{chi1}) and
(\ref{chi2}) suggests that numerical calculations are enough accurate. 
$\left[ \chi_{\bf p}\right]^2$ obtained this way is plotted in FIG. 2. The same
quantity was calculated in \cite{kaplan} using a different method. The 
agreement between FIG. 2 and the results of \cite{kaplan} is quite
satisfactory. 

Having calculated $\left[ \chi_{\bf p}^\Lambda \right]^2\left( \approx 
\left[ \chi_{\bf p}\right]^2\right)$ it is not difficult to
calculate the sum $\tilde G_E^\Lambda \left( \approx \tilde G_E\right)$ of all
finite regularised (unregularised) diagrams contributing to 
$G_E^\Lambda \ \left( G_E\right)$.  

Phase shifts corresponding to subtractive renormalization procedure 
can be calculated performing following steps:

1. Calculate $A_\pi$, calculate $A^\Lambda $ for very large $\Lambda $
   ($\Lambda\sim 10^5 {\rm MeV}$ has been taken);

2. Calculate 
\be
W(p)=-{1\over \tilde C}+\tilde G_E^\Lambda =
{\left[ \chi_{\bf p}^\Lambda\right]^2 \over A^\Lambda -A_\pi}
-I^\Lambda (p)-I_\pi^\Lambda (p) 
\label{w}
\ee
where $I^\Lambda (p)$ and $I_\pi^\Lambda (p)$ correspond to the first and second
divergent diagrams contributing to $G_E^\Lambda $ respectively. 

3. Calculate two divergent diagrams contributing to $G_E$ analytically, subtract
the second logarithmically 
divergent diagram at $p^2=-\mu^2$ and the first
linearly divergent diagram at $p^2=-\nu^2 (=-\kappa \mu^2)$ (one is free to
subtract these two diagrams at 
different points). In calculations of this work $\mu =240 \ {\rm MeV}$ and $\nu
=800 \ {\rm MeV}$ were taken 
($\kappa =100/9$). 
Note that the
large value of the second subtraction point (800 MeV) does not affect the power
counting.   

Subtracted diagrams give:

\be
I^R_\pi (p)={g_A^2\over 2f^2}\left( {m_\pi M\over 4\pi}\right)^2\left[ 
i\tan^{-1}\left( {2p\over m_\pi }\right) -{1\over 2}\ln \left( 1+{4p^2\over
m_\pi^2}\right) +\ln\left( 1+{2\mu\over m_\pi}\right)\right]
\label{IRpi}
\ee
and
\be
I^R(p)={M(\nu +ip)\over 4\pi }
\label{IR}
\ee

4. Calculate renormalized amplitude:
\be
A^R=A_\pi +{\left[ \chi_{\bf p}^\Lambda\right]^2\over W(p)+I^R(p)+I^R_\pi (p)} 
\label{AR}
\ee

5. Fit $C$ so as to reproduce phase shifts for very low energies ($p\sim 5 \
{\rm 
MeV}$). One gets
$C=-1/(17348 \ {\rm MeV}^2)$.   

6. Calculate phase shifts. Comparing phase shifts calculated for $\Lambda
=1\times 10^5 \ {\rm MeV}, 
2\times 10^5 \ {\rm MeV},
3\times 10^5 \ {\rm MeV}, 4\times 10^5 \ {\rm MeV}$ one sees that $A^R$ is
practically cut-off independent for very large $\Lambda $. 

These leading order phase shifts are plotted in FIG. 3.

As was argued in \cite{kaplan} $C_2$ coupling constant of the part of 
the next-to-leading order potential 
\be
V_1=C_2\left( {\bf p}^2+{\bf p'}^2\right)/2
\label{v1}
\ee
is larger than assumed by Weinberg's
power counting. This suggests that contributions of this part of 
the next-to-leading order potential are larger than of
others. This observation simplifies the actual next-to-leading order
calculations significantly. It is possible  
to sum up all diagrams obtained by iterating Lippman-Schwinger equation for the
potential $V=V_0+V_1$, where $V_0$ and $V_1$ are given by (\ref{potential}) and
(\ref{v1}) respectively. The result obtained in \cite{kaplan} using dimensional
regularization reads:
\be
A_1({\bf p})=A_\pi -{\left[ \chi_{\bf p}\right]^2\over \left[ \tilde C
-\alpha_\pi m_\pi MC_2+C_2{\bf p}^2\right]^{-1}-G_E}
\label{A1}
\ee 

As $\chi_{\bf p}$ and $\tilde G_E$ are already calculated
(although these quantities were calculated using cut-off regularization, as was
mentioned above 
the effect of finite cut-off is extremally small and hence
negligible), it is
straightforward to apply the same subtraction scheme to (\ref{A1}), fix $C_2$
so as to fit the data for $p=10 \ {\rm MeV}$
and calculate the phase shifts. The results are plotted in FIG. 3.
One can see that agreement with Nijmegen phase shift data \cite{nijmegen}
is quite
good. Obtained value 
$C_2=1/(8.5\times 10^9 \ {\rm MeV}^4)$ 
is indeed larger than assumed by Weinberg's counting.

One also can include $V_1$ perturbatively expanding (\ref{A1}) in $C_2$ and
retaining 
only first two terms. Calculated phase shifts are also plotted in FIG. 3. 
Comparing the phase shifts for perturbative and non-perturbative inclusion of
$V_1$ one sees that deviation is small for momenta up to 300 MeV. This suggests
that higher order contributions of $V_1$ are small as is expected from power
counting arguments. 

On the other hand it is straightforward to include pions perturbatively (as was
done in \cite{Gegelia}) following ideas of KSW counting. Using Those values for
$C$ and $C_2$ which were obtained above one gets {\it negative} phase shifts.
this suggests that the higher order contributions of the one pion exchange
potential are by no means small. Note that subtraction scheme used here is
different from the one considered in \cite{Gegelia}. In \cite{Gegelia} finite
diagrams were subtracted as well.
Repeating the calculations of this work with non-perturbatively included one
pion exchange potential using PDS scheme and substituting the
``best fit'' parameters from \cite{kaplan2} one gets again {\it negative} phase
shifts.   
This comparison of the phase shifts calculated using perturbative and
non-perturbative inclusion of pions supports the conclusion of \cite{Gegelia}
that the consistency of the perturbative
inclusion of pions in EFT approach to the $NN$-scattering problem is
questionable.

\medskip
\medskip
\medskip

{\bf ACKNOWLEDGEMENTS}


This work was carried out whilst the author was a recipient of an 
 Overseas Postgraduate
Research Scholarship and a Flinders University Research Scholarship
at the Flinders University of South Australia.


\end{document}